\documentclass[12pt]{article}
\usepackage[english]{babel}
\usepackage{a4}
\usepackage{graphicx}
\usepackage{amsfonts}
\usepackage{makeidx}
\begin{document}

\phantom{.}

\vspace*{1.cm}

\centerline{\large When teaching:}

\vspace*{1.cm}

\centerline{\Huge Out with magnitudes,}

\vspace*{0.7cm}

\centerline{\Huge in with monochromatic luminosities!}

\vspace*{1.5cm}

\centerline{\large Frank Verbunt}

\vspace*{0.5cm}

\centerline{\large Astronomical Institute University Utrecht, the Netherlands}

\vspace*{0.5cm}

\centerline{\large email: verbunt at astro.uu.nl}

\vspace*{2.5cm}

\section*{Abstract}

The goal of this document is to illustrate that teaching the
concepts of magnitudes is a needless complication in introductory
astronomy courses, and that use of monochromatic luminosities, rather than
arbitrarily defined magnitudes, leads to a large gain in
transparency. This illustration is done through three examples: the
Hertzsprung-Russell diagram, the cosmic distance
ladder, and interstellar reddening. I provide
conversion equations from the magnitude-based to the luminosity-based
system; a brief discussion; and a reference to sample lecture notes.

I suggest that we, astronomers in the 21st century,
{\bf abolish magnitudes and instead
use (apparent) monochromatic luminosities} in
non-specialist teaching. Given the large gain in transparency I
further propose that we seriously consider using (apparent)
monochromatic luminosities also in research papers, bringing
optical astronomy in line with astronomy at other wavelengths.

I would appreciate your opinion, comments and suggestions (see Sect.\,5). 
 
\section{The Hertzsprung-Russell diagram}

\begin{figure}
\centerline{\includegraphics[angle=270,width=0.55\columnwidth]{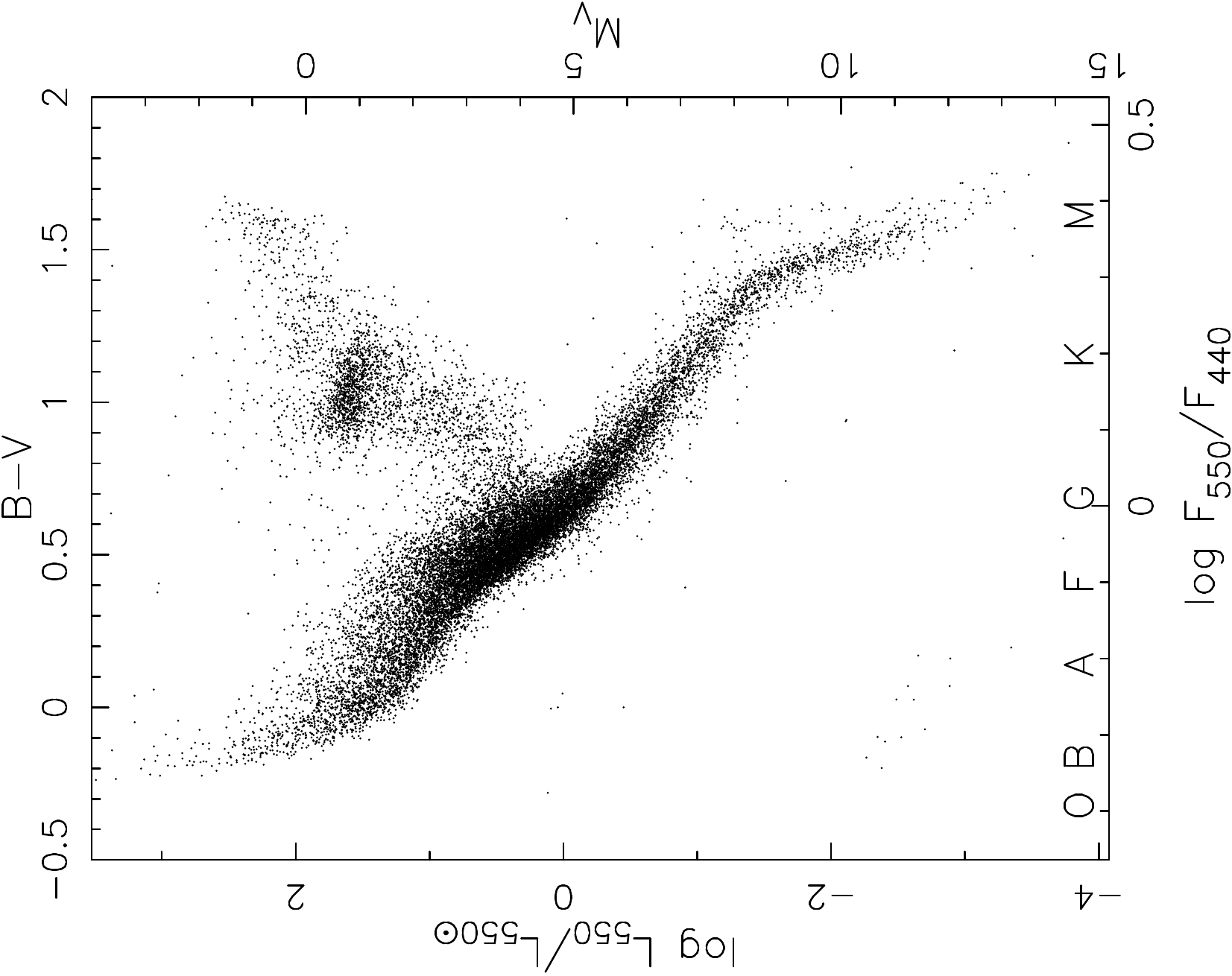}}

\caption{\it Monochromatic luminosity $L_{550}$ as a 
function of colour $F_{550}/F_{440}$ of 20305 stars with an
Hipparcos distance more accurate than 
10\%. The corresponding $M_V$ and $B-V$ are shown at the
right hand side and top. \label{f:newhr}}
\end{figure}

In analogy with the total or bolometric luminosity, we define the
{\em monochromatic luminosity} $L_\lambda$ as
\begin{equation}
L_\lambda = 4\pi R^2 F_\lambda = 4\pi d^2 f_\lambda
\label{e:monlumin}\end{equation}
Here $F_\lambda$ and $f_\lambda$ are fluxes
per unit wavelength, i.e.\ monochromatic fluxes, at the
stellar surface and at Earth, respectively, $R$ and $d$ are the
radius of and distance to the star.
For the Sun, the flux at 550\,nm in a band of 1\,nm is
$f_{550}\simeq2\,\mathrm{W\,m}^{-2}\,\mathrm{nm}^{-1}$, and thus
with $d=1$\,AU, $L_{550\odot}\simeq5.6\times10^{23}\,\mathrm{W\,nm}^{-1}$.
(In fact, this flux is averaged over the Johnson V filter, and one may
prefer to denote it $f_V$.)

The Hipparcos catalogue provides distances $d$ and monochromatic
fluxes $f_{550},f_{440}$ which with Eq.\,\ref{e:monlumin}
give the monochromatic luminosity $L_{550}$ and colour
$F_{550}/F_{440}=f_{550}/f_{440}$. (For the moment we
assume that interstellar absorption is negligible.)  We combine these
in a colour - luminosity diagram Figure\,\ref{f:newhr}. This Figure
tells us immediately that a B main-sequence star emits about 100 times
more photons at 550\,nm than the Sun, and a K dwarf about 10 times
fewer. It also tells us directly that an early G star emits about as
much flux at 550\,nm as at 440\,nm.

Such information is also present in the diagram labeled with
$M_V$ and $B-V$, but very much hidden. 

\begin{figure}
\centerline{\includegraphics[width=0.5\columnwidth]{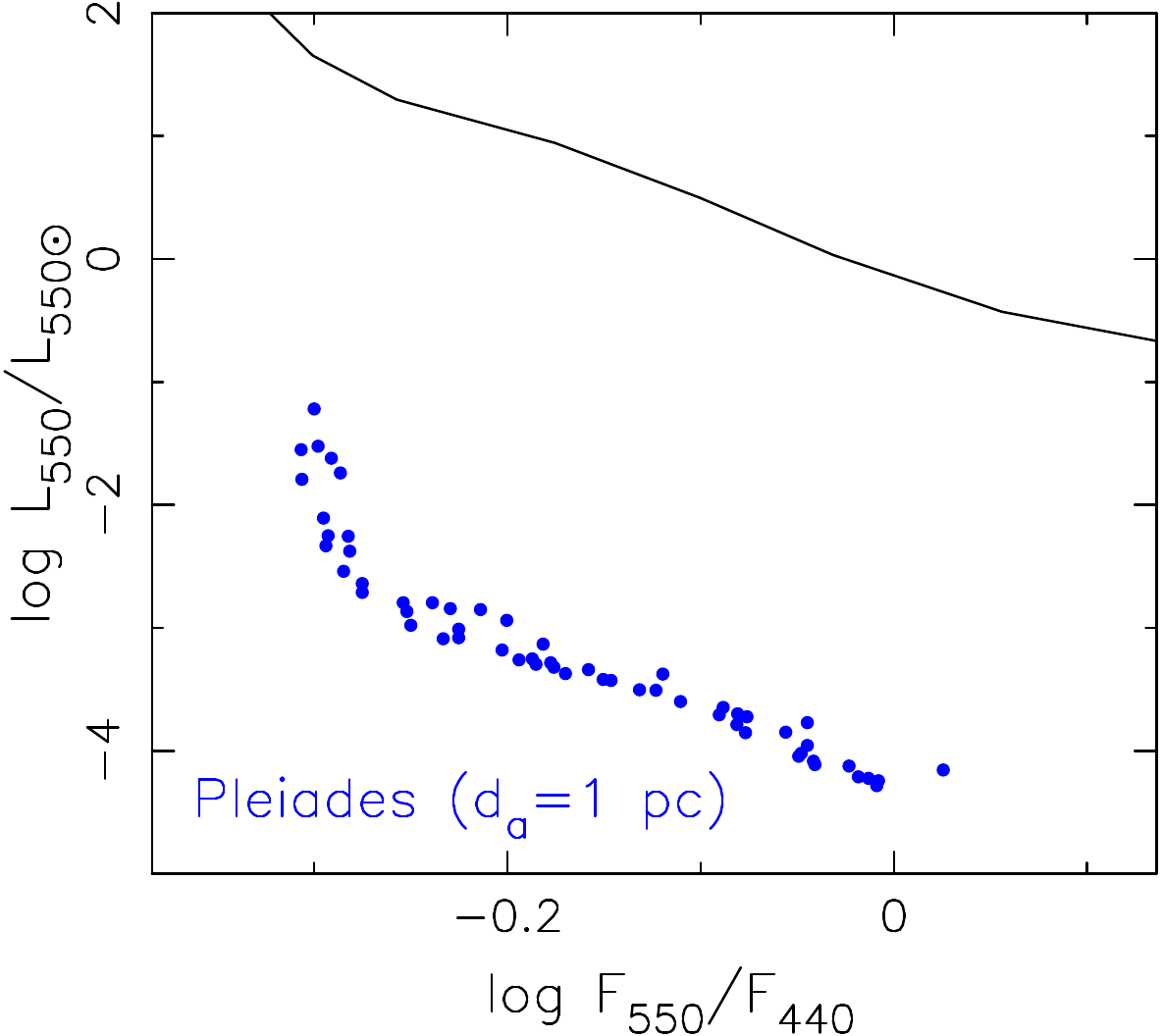}}

\caption{\it The lower envelope of the main
sequence as found with Hipparcos (solid line, see Figure\,\ref{f:newhr}) and
{\em apparent} luminosities of stars in the Pleiades ($\bullet$),
for an assumed distance $d_a=1$\,pc.
\label{f:pleiades}}
\end{figure}

\section{The cosmic distance ladder}

For distances beyond the reach of parallax measurements, several
methods are available that are based on monochromatic
luminosities. For these we define the {\em apparent monochromatic
luminosity} $L_{\lambda a}$, based on an {\em assumed distance}
$d_a$, as:
\begin{equation}
L_{\lambda a} = 4\pi {d_a}^2 f_\lambda
\label{e:amonlumin}\end{equation}
We allow any choice for $d_a$, which means that {\em the apparent
luminosity must be given in tandem with an assumed distance}.
The logarithms of the real and apparent monochromatic luminosities
differ, with Eqs.\,\ref{e:monlumin} and \ref{e:amonlumin}, by:
\begin{eqnarray}
\Delta L & \equiv & \log L_{\lambda} - \log  L_{\lambda a} 
=  2 \log {d\over d_a} \nonumber \\
& = & \log {L_{\lambda}\over L_{\lambda\odot}} - \log {L_{\lambda a}\over L_{\lambda\odot}}
\label{e:deltal}\end{eqnarray}
When we can determine this shift, the distance follows as
\begin{equation}
d = 10^{0.5\Delta L}d_a
\label{e:dism}\end{equation}

\subsection{Main-sequence fitting}

As an example I show in Figure\,\ref{f:pleiades} the apparent
luminosity at 550\,nm for stars in the Pleiades for an assumed distance $d_a=1$\,pc,
together with the lower envelope of the main-sequence for nearby
stars, from Figure\,\ref{f:newhr}.  A vertical shift of $\Delta
L\simeq4.4$ will put the main sequence of the Pleiades just above the
lower envelope for nearby stars.  The distance of the Pleiades follows
with Eq.\,\ref{e:dism} as about 160\,pc.

\subsection{The Cepheid period-luminosity relation}

\begin{figure}
\centerline{\includegraphics[width=\columnwidth]{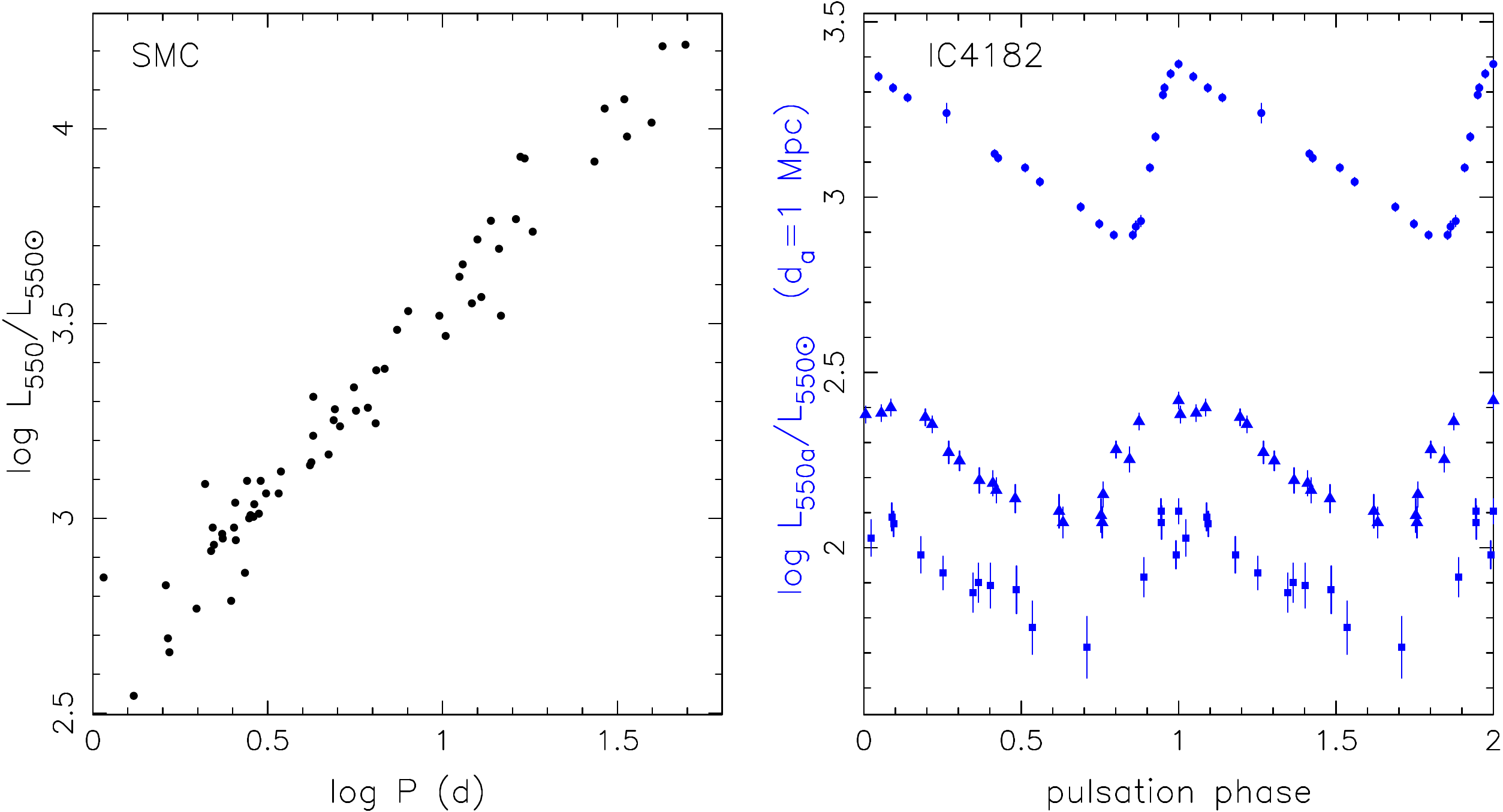}}

\caption{\it The period-luminosity relation for Cepheids in the
Small Magellanic Cloud (left frame), and apparent monochromatic luminosities
$L_{550a}$ for assumed distance $d_a=1$\,Mpc for three Cepheids in
IC\,4182. The periods of the Cepheids in IC\,4182 are for decreasing
luminosity: 42\,d, 9.24\,d and 4.26\,d. \label{f:cepheid}}
\end{figure}

Figure\,\ref{f:cepheid} shows the period - luminosity relation for
Cepheids in the Small Magellanic Cloud (left frame; the luminosity
shown for each Cepheid is the average of the maximum and minimum value
of $\log L_{550}$\footnote{This requires a distance for the SMC; in my
lecture notes I determine the distance to h and $\chi$ Persei by
main-sequence fitting, and the distance to the SMC by demanding that
the period-luminosity relation passes through the values for the
Cepheids in h and $\chi$ Persei}) and lightcurves of three Cepheids in
the galaxy IC\,4182 (right frame; apparent luminosities for an assumed
distance $d_a=1$\,Mpc). The brightest Cepheid in IC\,4182 shown has a
period of 42\,d and an average apparent luminosity $\log
L_{550a}/L_{550\odot}\simeq3.15$.  From the left frame of
Figure\,\ref{f:cepheid} we see that the correct luminosity at 42\,d
is $\log L_{550}/L_{550\odot}\simeq4.2$, i.e.\ $\Delta
L\simeq1.05$. With Eq.\,\ref{e:dism} a distance of about 3.3\,Mpc follows
for IC\,4182.

\subsection{Maximum luminosity of supernova Ia}

\begin{figure}
\centerline{\includegraphics[angle=270,width=0.5\columnwidth]{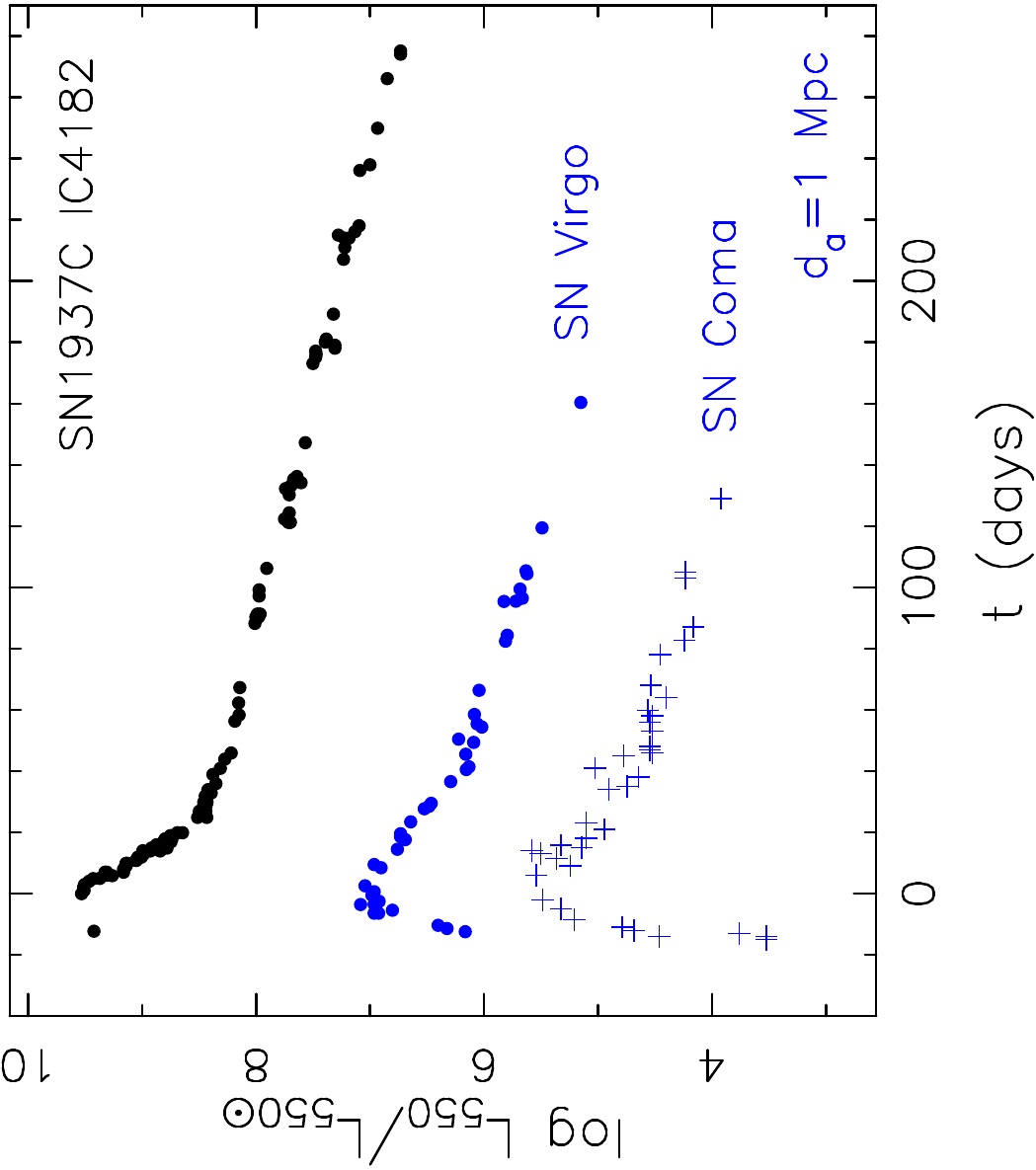}}

\caption{\it The lightcurve at 550\,nm for supernova of type Ia 1937C
in IC\,4182, and lightcurves of apparent monochromatic luminosities 
for supernovae of type Ia in the Virgo and Coma clusters
(computed for $d_a=1$\,Mpc, the lightcurve for Coma is a composite of 
several supernovae).
\label{f:sn1a}}
\end{figure}

In 1937 a supernova of type Ia was observed in IC\,4182. With the
distance of this galaxy derived from the Cepheids in it, we can
compute the monochromatic luminosities of this supernova. 
The resulting lightcurve is shown in 
Figure\,\ref{f:sn1a}.  In Figure\,\ref{f:sn1a} I also show the
lightcurve for a supernova Ia in the Virgo cluster, and a composite
lightcurve of several supernovae Ia in the Coma cluster, computing
apparent luminosities for an assumed distance $d_a=1$\,Mpc.  We see
from the maxima (at $t=0$\,d) that $\Delta L\simeq2.5$ for Virgo and
$\Delta L\simeq 4$ for Coma, giving distances with Eq.\,\ref{e:dism}
of about 18\,Mpc and 100\,Mpc, respectively.

\section{Interstellar absorption and reddening}

Absorption and scattering of radiation by interstellar matter
is characterized by the optical depth $\tau_\lambda$, where the
subscript indicates the wavelength dependence. The flux $f_\lambda^c$
corrected for the absorption is given by
\begin{equation}
f_\lambda^c=f_\lambda e^{\tau_\lambda}
\label{e:absflux}\end{equation}
{\em It is this corrected flux which should be entered in 
Eqs.}\,\ref{e:monlumin},\ref{e:amonlumin}, when $\tau_\lambda>0$.

\begin{figure}
\includegraphics[width=\columnwidth]{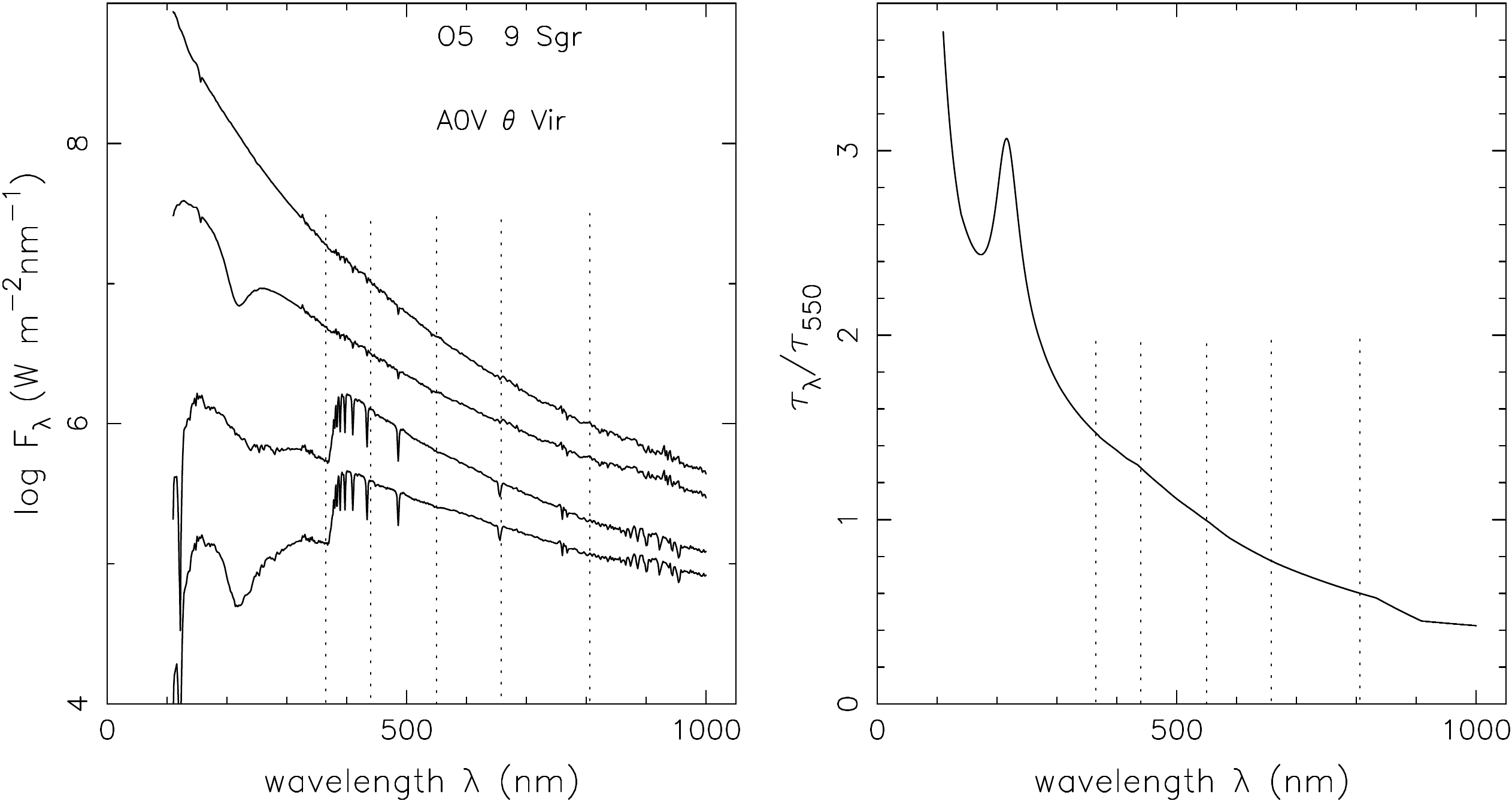}

\caption[o]{\it Left: for two spectral types the figure shows the
standard spectrum and below it the spectrum absorbed by interstellar
matter (with $\tau_{550}=1$). Right: interstellar absorption as a
function of wavelength, normalized on the absorption at 550\,nm.  The
overall smooth decline towards longer wavelength is due to absorption
by gas, the extra opacity near 220\,nm is due to dust.
\label{newred}}
\end{figure}

Because of the wavelength dependence of the absorption, the colours
of the star change. With Eq.\,\ref{e:absflux} we can relate the
change in colour to the optical depths at the different wavelengths.
For the visual-to-blue flux ratio, we find
\begin{equation}
\log {f_{550}^c\over f_{440}^c}=\log {f_{550}\over f_{440}}
+\left(\tau_{550}-\tau_{440}\right)\log e
\label{e:redcol}\end{equation}
Generalizing this equation to arbitrary $\lambda$, we see that by
taking the ratio of the intrinsic colour $f_\lambda^c/f_{550}^c$ and
the observed colour $f_\lambda/f_{550}$ we can derive the ratio
$\tau_\lambda/\tau_{550}$. Figure\,\ref{newred} shows how this
works. From the line strengths we can derive the spectral types of the
stars as O5 and A0\,V, even if the spectrum is affected by
interstellar absorption. The dependence of $\tau_\lambda$
on wavelength is found by dividing the observed, absorbed spectrum
by the correct, standard spectrum.
The result is shown in Figure\,\ref{newred} (right).

If we know the intrinsic colour of the star, we can
derive the interstellar absorption even if we do not know the
distance. For example, from the
data used to make Figure\,\ref{newred} we have
$\tau_{440}\simeq1.32\,\tau_{550}$. Hence
\begin{equation}
\tau_{550}\simeq 3.1 \left(\tau_{440}-\tau_{550}\right)
\end{equation}
where the right hand side is found from Eq.\,\ref{e:redcol}.
\begin{figure}
\includegraphics[width=\columnwidth]{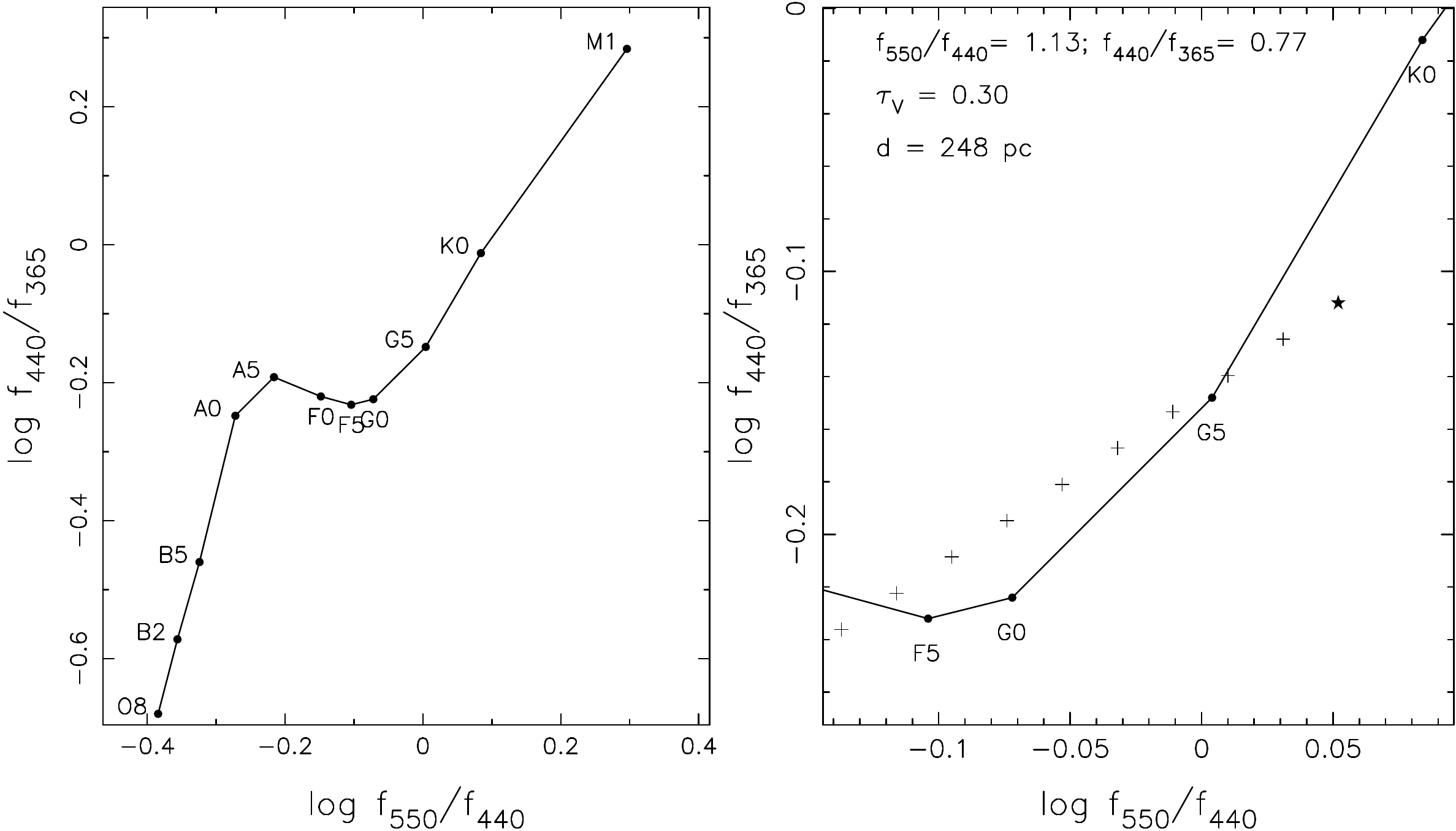}

\caption[o]{\it Left: colour-colour diagram for main-sequence stars.
Right: enlarged region of the plot, with the observed colours
of a star (*) and the colours (+) after correction for absorption
with $\tau_{550}=0.15,0.30,0.45\ldots$. 
\label{newcoldis}}
\end{figure}

If we have measurements at three wavelengths, i.e.\ two independent
colours, we can also determine the reddening, as illustrated in
Figure\,\ref{newcoldis}. The left frame shows the main sequence
colours in the absence of reddening, the right frame shows a reddened
star that due to reddening no longer lies on this
sequence. Corrections to the colours have been applied for assumed
values of $\tau_{550}$ in steps of 0.15. It is seen that a correction
for $\tau_{550}=0.3$ puts the star on the main-sequence colour-colour
relation, close to a G5\,V star. With the intrinsic luminosity
$L_{550}=L_{550\odot}$, the observed flux $f_{550}=5.65\times10^{-16}$
W\,s$^{-1}$\,nm$^{-1}$, and the determined reddening $\tau_{550}=0.3$
the distance of the star can be determined.  The Figure also shows
that the solution is not unique: an early F star reddened by
$\tau_{550}\simeq1.2$, or an even more reddened B star, is also
possible. Often the use of other colours allow selection of a unique
solution.

\section{Conversion formulae}

Absolute visual magnitudes to luminosity at 550\,nm:
\begin{equation}
\log {L_{550}\over L_{550\odot}} = 0.4\left[M_{V\odot}-M_V\right]=
0.4\left[4.81-M_V\right]
\label{e:mvtolv}\end{equation}

\noindent Colour to flux ratio (for $\tau_\lambda=0$):
\begin{eqnarray}
\log{F_{550}\over F_{440}} & = & \log{f_{550}\over f_{440}} 
=  0.4\left[B-V-0.66\right] \\
\log{F_{806}\over F_{550}} & = & \log{f_{806}\over f_{550}} 
=  0.4\left[V-I-1.35\right] \\
\log{F_{440}\over F_{365}} & = & \log{f_{440}\over f_{365}} 
=  0.4\left[U-B-0.54\right] 
\label{e:fvofb}\end{eqnarray}
The constants follow from the definitions of zero magnitude
for $U$, $B$, $V$, and $I$.

\noindent Apparent visual magnitudes to monochromatic luminosity:
\begin{equation}
\log{L_{550}\over L_{550\odot}} =2 \log {d\over\mathrm{1\,pc}} -0.4[V+0.19]
\label{e:vdtolv}\end{equation}

\noindent Absorption in magnitudes to optical depth:
\begin{equation}
\tau_\lambda = {0.4 A_\lambda\over\log e}
\end{equation}
hence colour excess to difference in optical depth:
\begin{equation}
\tau_{440}-\tau_{550} = {0.4E(B-V)\over\log e}
\end{equation}

\section{Discussion}

{\bf What is wrong with magnitudes?}
The definition of the absolute magnitude as used in the original form of
the colour-magnitude diagram 
$$ M_\lambda = m_\lambda + 5 - 5\log{d\over\mathrm{1\,pc}}=
 -2.5\log f_\lambda + c_\lambda - 5\log{d\over\mathrm{10\,pc}} $$
has no less than four arbitrary elements,
which we have to explain to students:
\begin{enumerate}
\item the minus-sign, which causes high magnitudes to correspond to low
fluxes
\item the factor 2.5
\item the flux at zero apparent magnitude $m_\lambda$, as expressed in
$c_\lambda$ (which has to be defined for each filter -- or $\lambda$
-- in each filter system!)
\item the distance of 10\,pc 
\end{enumerate}
Monochromatic luminosity has no arbitrariness.
I have no quarrel with Claudius Ptolemaios for his assigning
magnitudes between 1 and 7 to the stars that are
visible to the naked eye, but I somehow think that
he will forgive us if we switch to a more convenient notation.

{\bf Apparent luminosities.} 
The advantage of leaving the assumed distance $d_a$ free, is that
a suitable value can be chosen for each application. When determining
the distance to the Coma cluster a reference distance of 10\,pc
is less suitable than a reference distance measured in Mpc. The disadvantage
is that each apparent luminosity must be accompanied explicitly by
a mention of the assumed distance $d_a$.
This disadvantage can be removed if we agree to limit choices
of $d_a$ to integer powers of 10. The power used can then be
added to the index: $L_{550a6}$ would denote the apparent
luminosity for $d_a=1$\,Mpc, and $L_{550a1}$ the apparent
luminosity for $d_a=10$\,pc. The numbers 1-9 would suffice for
any practical purpose.

In practice a flux often represents a weighted average over
a filter. One can then replace the wavelength in the index with the
filter name. Thus $L_{Ua4}$ is the apparent luminosity
derived from the average flux in the $U$ filter, for $d_a=10$\,kpc.

{\bf Sample lecture notes.} To test the viability of using monochromatic
luminosities, I have re-written sections of my lecture notes for a course on
{\em Life of the stars: from dust to black holes} that I teach to
first year physics \&\ astronomy students in Utrecht. The course is in Dutch
and uses SI units. English versions of the re-written sections can be found
on my website:

\centerline{{\tt www.astro.uu.nl/}$\sim${\tt verbunt/onderwijs/intro/new}}

\noindent
as sample lecture notes that use (apparent) monochromatic luminosities.
More Tables and Figures can be found in these.

{\bf Some final remarks.}
If you have already found a similar, indeed better, way to replace magnitudes,
please let me know.

It may be argued that an advantage of the magnitude system is that it
makes explicit that any flux measurement is based on flux ratios with
calibrated sources.  It appears to me that this should be obvious
anyway: remember that radio and X/$\gamma$ astronomers do very well
without magnitudes.

If a miracle happens, i.e.\ if reason prevails, and enough astronomers
agree that we should stop using magnitudes, we may wish to avoid a
proliferation of different notations and choices by asking the
appropriate committees of the {\em International Astronomical Union}
to define the method and notation of preference.

If enough of you encourage me, I will ask the IAU to do so.

\vspace*{3.5cm}

\noindent
{\small {\bf data used in the figures}: (1) Perryman et al.\ 1995 A\&A
304, 69.  (2) Johnson et al.\ 1952 ApJ 117, 313. (3) Arp 1960 AJ 65,
404; Saha et al.\ 1994 ApJ 425, 14. (4) Baade \&\ Zwicky 1938, ApJ 88,
411 (1937C); Kimeridze \&\ Tsvetkov 1987 Soviet Astr.\ 25, 513; Barbon
et al.\ 1989 A\&A 220, 83 (1984A in NGC4419); Zwicky 1961 PASP 73, 185
(1961D), Zwicky \&\ Barbon 1967 AJ 72, 1366 (1962A), Barbon 1978 AJ
83, 13 (1963F,M), Kohoutek \&\ Kowal 1978 PASP 90, 565 (1975F) (5)
Gunn \&\ Stryker 1983 ApJS 52, 121, Seaton 1979 MNRAS 187, 75P (6)
Bessell et al.\ 1998 A\&A 333, 231}

\end{document}